\documentclass{article}

\usepackage{amssymb}
\usepackage{amsthm}
\usepackage{amscd}

\newtheorem{thm}{Theorem}[section]
\newtheorem{prop}[thm]{Proposition}
\newtheorem{lem}[thm]{Lemma}
\newtheorem{cor}[thm]{Corollary}

\theoremstyle{definition}
\newtheorem{dfn}[thm]{Definition}

\theoremstyle{remark}
\newtheorem*{rem}{Remark}

\newcommand{\C}{\mathbb{C}}
\newcommand{\T}{\mathbb{T}}

\newcommand{\A}{\mathcal{A}}
\newcommand{\B}{\mathcal{B}}
\newcommand{\sC}{\mathcal{C}}

\newcommand{\G}{\mathcal{G}}
\newcommand{\sT}{\mathcal{T}}
\newcommand{\Tr}{\mathrm{Tr}}
\newcommand{\Id}{\mathrm{Id}}
\newcommand{\hq}{\widehat{q}}
\renewcommand{\S}{\mathcal{S}}

\def\h#1{ \widehat{#1} }
\def\M#1{ \mathrm{Map}({#1},G) }

\title{Gerbes in classical Chern-Simons theory}

\author{Kiyonori Gomi
\thanks{The author's research is supported by Research Fellowship of the Japan Society for the Promotion of Science for Young Scientists.}}

\date{}

\begin{document}

\maketitle

\begin{abstract}
We construct geometrically a gerbe assigned to a connection on a principal $SU(2)$-bundle over a closed oriented 1-dimensional manifold. If the connection is given by the restriction of a connection on a bundle over a compact 2-manifold bounding the 1-manifold, then we have a natural object in the gerbe. The gerbes and the objects satisfy certain fundamental properties, e.g.\ gluing law. 
\end{abstract}


\section{Introduction}
It is well known that the value of the Chern-Simons action functional (CS action) \cite{F1, R-S-W, W} of a connection $A$ on a principal $SU(2)$-bundle $P$ over a closed oriented 3-manifold $M$ is well-defined as a complex number of unit norm: $e^{2\pi\sqrt{-1} S_M(A)} \in \T$. For a principal bundle $Q$ over a closed oriented 2-manifold $\Sigma$, there exists a Hermitian line bundle over the space of connections on $Q$. We call the line bundle Chern-Simons line bundle and denote the unit norm elements of the fiber at a connection $a$ by $T_Q(a)$. This set admits a simply transitive action of $\T$ and is so-called a \textit{principal $\T$-space}, or a \textit{$\T$-torsor}. If the 3-manifold $M$ has boundary, then the value of the CS action depends on a certain boundary condition. Hence the CS action is defined as $e^{2\pi\sqrt{-1} S_M(A)} \in T_{\partial P}(\partial A)$. The CS action and the CS line bundle (or $\T$-torsor $T_Q(a)$) satisfy certain fundamental properties \cite{F1, F2} which are the classical counterpart of the axioms of the Topological Quantum Field Theory \cite{A}. In particular, we can compute the CS action with $\partial M = \emptyset$ from the CS action with $\partial M \neq \emptyset$.

For a 2-manifold with boundary we cannot define the CS line bundle without any choice of a certain boundary condition \cite{F1}. This is similar to the definition of the CS action for a 3-manifold with boundary. According to Freed \cite{F2}, these phenomena arise in all dimensions and are described by using higher analogue of principal homogeneous spaces. For a closed 2-manifold we already have the $\T$-torsor $T_Q(a)$ assigned to each connection on $Q$. For a closed 1-manifold a sort of category called a \textit{$\T$-gerbe} \cite{F2} appears. Let $\sT$ be the category of $\T$-torsors. The category has an abelian group-like structure (Proposition \ref{prop_agls_torsors}). A category on which $\sT$ acts simply transitively in the categorical sense is called a $\T$-gerbe (see Section \ref{la_intro_gerbe}). 

In the case of Chern-Simons theory with a finite structure group, the $\T$-gerbes associated with closed 1-manifolds are given by a certain integration theory of cocycles \cite{F2}. In the case of Chern-Simons theory with a continuous structure group, a construction of the $\T$-gerbes is proposed extending the integration theory \cite{F3, F4}. The purpose of this paper is a geometric construction of the $\T$-gerbes. We only consider the case that the structure group $G$ is $SU(2)$, since we can easily generalize the construction to the case that $G$ is a simple, simply connected, connected, compact Lie group.

Here we briefly explain the construction. Let $S$ be a closed oriented 1-manifold (disjoint union of unparameterized circles), and $R$ a principal bundle over $S$ whose structure group is $G = SU(2)$. We denote the space of connections on $R$ by $\A_R$, the space of sections of $R$ by $\S_R$, and the loop group \cite{P-S} by $\G_S = \M{S}$. Because $\G_S$ acts on $\S_R$ by the pointwise product, we have a (trivial) $\G_S$-bundle $B_R = \A_R \times \S_R$ on $\A_R$. We denote by $B_R(a)$ the fiber at $a \in \A_R$. Obviously $B_R(a)$ admits a simply transitive action of $\G_S$, and is a $\G_S$-torsor. Using the method of Mickelsson \cite{Mi} we can construct a central extension of $\G_S$ by $\T$
$$
\begin{CD}
1 @>>> \T @>>> \h{\G}_S @>>> \G_S @>>> 1.
\end{CD}
$$
For this central extension a \textit{lifting} of $B_R(a)$ is a pair $(\h{B}, \hq)$, where $\h{B}$ is a manifold with a simply transitive $\h{\G}_S$-action (i.e.\ $\h{\G}_S$-torsor) and $\hq : \h{B} \to B_R(a)$ is an equivariant map under the actions of the groups. The category $\B_R(a)$ of liftings of $B_R(a)$ admits an action of $\sT$ and we have a $\T$-gerbe assigned to each connection $a \in \A_R$. We can regard $\B_R(a)$ as the category of liftings of the structure group of the bundle $B_R(a) \to \{a\}$ to the central extension. Moreover we construct a lifting $\h{B}_Q(A) \in \textrm{Obj}(\B_{\partial Q}(\partial A))$ assigned to a connection $A$ on a bundle $Q$ over a 2-manifold $\Sigma$ with boundary. As is also indicated in \cite{F2}, the $\T$-gerbes and the liftings satisfy the fundamental properties (Theorem \ref{thm_axiom_CSgerbe}) such as gluing law.

We remark that the description of gerbes in this paper is \textit{local}. As the CS line bundle corresponding to $T_Q(a)$ is defined over $\A_Q$, there exists a global geometric object over $\A_R$ corresponding to $\B_R(a)$. Such a geometric object is described as a \textit{Dixmier-Douady sheaf of groupoids} or a \textit{DD gerbe} \cite{Br1,Br2}, and will be observed in the forthcoming paper \cite{Go}.

\medskip

This paper is organized as follows. In Section \ref{la_CSaction_CSlinebundle} we recall the CS action and the CS line bundle. In Section \ref{la_WZ_ce} we study the Wess-Zumino term and the construction of the central extensions of loop groups. The construction is essential to the main properties of the $\T$-gerbes and the liftings. In Section \ref{la_intro_gerbe} we introduce $\T$-torsors and $\T$-gerbes. In Section \ref{la_lifting_gerbe} we define a $\T$-gerbe associated with liftings. We observe the correspondence between the operations on liftings and that on $\T$-gerbes. In Section \ref{la_results} we state the main results of this paper. We define the $\T$-gerbe $\B_R(a)$ and the lifting $\h{B}_Q(A)$. The fundamental properties are proved here.


\section{The Chern-Simons action functional and the Chern-Simons line bundle}
\label{la_CSaction_CSlinebundle}

We recall the Chern-Simons action and the Chern-Simons line bundle \cite{F1, R-S-W, W}. 

\smallskip

In this paper $G$ represents $SU(2)$ otherwise stated. Let $M$ be a compact oriented smooth 3-manifold, and $\pi : P \to M$ a smooth principal $G$-bundle. We denote the space of smooth connections on $P$ by $\A_P$ and the space of smooth sections of $P$ by $\S_P$. Because any $SU(2)$-bundle over a compact manifold whose dimension is less than 4 is topologically trivial, $\S_P$ is not empty. The space of smooth maps $\G_M = \M{M}$ is a Lie group by the pointwise product and acts on $\S_P$ by $s \mapsto s \cdot g$. We fix an integer $k$ throughout this paper.

\begin{lem}
For $A \in \A_P$ and $s \in \S_P$ we put
\begin{eqnarray}
\tilde{S}_M(A,s) = \frac{k}{8\pi^2} \int_M 
s^* \Tr \left( A \wedge dA + \frac{2}{3} A \wedge A \wedge A \right) .
\end{eqnarray}
For $g \in \G_M$ we have
\begin{eqnarray} 
\tilde{S}_M(A,s \cdot g) - \tilde{S}_M(A,s) 
& = & 
\frac{k}{8\pi^2} \int_{\partial M} \Tr(s^*A \wedge dg g^{-1})
\label{difference_CS} \\
& - &
\frac{k}{24\pi^2} \int_M \Tr(g^{-1}dg \wedge g^{-1}dg \wedge g^{-1}dg) .
\nonumber
\end{eqnarray}
\end{lem}

This lemma is checked by a direct calculation. If a 3-manifold $M$ is compact and has no boundary (closed), then the first term of the right hand side of (\ref{difference_CS}) vanishes and the second term is an integer. So we have the following.

\begin{prop}
Let $M$ be a closed oriented smooth 3-manifold. If we put $\exp 2\pi \sqrt{-1} S_M(A) := \exp 2\pi \sqrt{-1} \tilde{S}_M (A,s)$ for some $s \in \S_P$, then the \textit{Chern-Simons action functional} (CS action) is well-defined as an element of the unit circle $\T = \{ z \in \C |\ |z| = 1 \}$
\begin{eqnarray}
e^{ 2\pi\sqrt{-1} S_M(A) } \in \T.
\end{eqnarray}
\end{prop}

In order to formulate the CS action in the case that a 3-manifold has boundary, we need a certain line bundle.

\begin{lem}
Let $\Sigma$ be a closed oriented smooth 2-manifold. 

(a) For $g \in \G_\Sigma$ we define the \textit{Wess-Zumino term (WZ term)} by
\begin{eqnarray}
e^{ W_\Sigma(g) } = \exp \frac{k\sqrt{-1}}{12\pi} 
\int_{\tilde{\Sigma}} \Tr( \tilde{g}^{-1} d\tilde{g} )^3 ,
\label{WZ_closed}
\end{eqnarray}
where $\tilde{\Sigma}$ is a compact 3-manifold bounding $\Sigma$ and $\tilde{g} \in \G_{\tilde{\Sigma}}$ is an extension of $g$. Then the WZ term satisfies the Polyakov-Wiegmann formula
\begin{eqnarray}
e^{ W_\Sigma(g_1 g_2) } & = &
\exp \Gamma_\Sigma(g_1,g_2) e^{ W_\Sigma(g_1) } e^{ W_\Sigma(g_2) } ,
\label{PW-formula_1} \\
\Gamma_\Sigma(g_1, g_2) & = &
-\frac{k\sqrt{-1}}{4\pi} \int_\Sigma 
\Tr( {g_1}^{-1}dg_1 \wedge dg_2 {g_2}^{-1}) .
\label{cocycle_2}
\end{eqnarray}

(b) For a smooth $G$-bundle $Q$ over $\Sigma$ we define a function $c_\Sigma : \A_Q \times \S_Q \times \G_\Sigma \to \T$ by
\begin{eqnarray}
c_\Sigma(A,s,g) = 
\exp \frac{k\sqrt{-1}}{4\pi} \int_\Sigma \Tr(s^*A \wedge dgg^{-1}) \cdot 
\left( e^{W_\Sigma(g)} \right)^{-1} . \label{cocycle_1}
\end{eqnarray}
The function satisfies the following formula
\begin{eqnarray}
c_\Sigma(A,s,g_1 g_2) = c_\Sigma(A,s \cdot g_1,g_2) c_\Sigma(A,s,g_1) .
\label{cocycle_condition_1}
\end{eqnarray}
\end{lem}

The WZ term (\ref{WZ_closed}) is independent of the choice of the bounding manifolds and the extensions. So it is well-defined as an element of $\T$. A direct calculation establishes the above lemma.

\begin{prop} \label{thm_cslinebundle}
For a closed oriented smooth 2-manifold $\Sigma$ we define a quotient space 
\begin{eqnarray}
L_{(Q,\Sigma)} = \A_Q \times \S_Q \times \C / \sim 
\end{eqnarray}
by $(A, s, z) \sim (A, s \cdot g, c_\Sigma(A,s,g) z)$ for $g \in \G_\Sigma$. If we define a projection $\Pi : L_{(Q,\Sigma)} \to \A_Q$ by $\Pi([A,s,z]) := A$, then we have a line bundle. Moreover the line bundle has a Hermitian structure defined by the standard one on $\C$.
\end{prop}

\begin{proof}
We can check the well-definedness of the equivalence relation using (\ref{cocycle_condition_1}). Because $\G_\Sigma$ freely acts on $\S_Q$, $L_{(Q, \Sigma)}$ is indeed a line bundle. The function $c_\Sigma$ is $\T$-valued, so the standard Hermitian structure on $\C$ defines that on $L_\Sigma$.
\end{proof}

We call the line bundle the \textit{Chern-Simons line bundle}. We denote the set of unit norm elements of the fiber of $L_{(Q, \Sigma)}$ at $a$ by $T_{(Q, \Sigma)}(a)$. It is clear that $T_{(Q, \Sigma)}(a)$ is a $\T$-torsor (see Definition \ref{dfn_torsor}). When the underlying manifold $\Sigma$ is obvious, we abbreviate $L_{(Q, \Sigma)}$ and $T_{(Q, \Sigma)}(a)$ as $L_Q$ and $T_Q(a)$.

\begin{prop} \label{thm_csaction_boundary}
Let $P \to M$ be a $G$-bundle over a compact oriented smooth 3-manifold with boundary. If we define the \textit{CS action} by $e^{ 2\pi\sqrt{-1} S_M(A) } := [ \partial A, \partial s, e^{ 2\pi\sqrt{-1} \tilde{S}_M(A,s)} ]$ for some $s \in \S_P$, then the CS action is well-defined as a unit norm element of the fiber of the CS line bundle 
\begin{eqnarray}
e^{ 2\pi\sqrt{-1} S_M(A) } \in T_{(\partial P, \partial M)}(\partial A).
\end{eqnarray}
\end{prop}

\begin{proof}
If we take the other section $s'$, then there exists a unique $g \in \G_M$ such that $s' = s \cdot g$. By (\ref{difference_CS}) we have
\begin{eqnarray*}
[\partial A, \partial s', e^{ 2\pi\sqrt{-1} \tilde{S}_M(A,s')} ] 
& = &
[\partial A, \partial s \cdot \partial g, 
c_{\partial M}(\partial A, \partial s, \partial g)
e^{ 2\pi\sqrt{-1} \tilde{S}_M(A,s)} ] \\
& = & 
[\partial A, \partial s, e^{ 2\pi\sqrt{-1} \tilde{S}_M(A,s)} ] .
\end{eqnarray*}
Hence the CS action is well-defined. Obviously it is of unit norm.
\end{proof}

The CS action and the CS line bundle are compatible with some operations on manifolds. These fundamental properties are summarized in \cite{F1}. We do not state here, but the same type of properties for the WZ term are described in the next section (Proposition \ref{thm_axioms_WZ}).


\section{The Wess-Zumino term and the central extensions of loop groups}
\label{la_WZ_ce}

We observe the definition of the WZ term in the case that the 2-manifold has boundary and the construction of the central extensions of loop groups. The properties of the WZ term and the central extensions play an essential role in the main result of this paper.

\smallskip

First we associate a disk to each closed 1-manifold.

\begin{dfn}
Let $S$ be a closed smooth 1-manifold.

(a) If $S$ is connected $(S \cong S^1)$, then we define a disk $D_S$ as the cone 
\begin{eqnarray*}
D_S = S \times [0,1] / S \times \{0\} , 
\end{eqnarray*}
i.e.\ the set obtained from $S \times [0,1]$ by collapsing $S \times \{0\}$ to a point. The disk $D_S$ has the topology coinduced by the projection $S \times [0,1] \to D_S$. The structure of smooth manifold is defined by taking $D_S$ itself as an open cover of $D_S$, and constructing a coordinate function $D_S \to D^2$ from an identification $S \to S^1$. If $S$ is oriented, then $D_S$ is oriented too.

(b) If $S$ is not connected, then we can express $S = S_1 \sqcup \cdots \sqcup S_n$ (disjoint union), where $S_i$ is a circle. In this case we put $D_S := D_{S_1} \sqcup \cdots \sqcup D_{S_n}$.
\end{dfn}

\begin{lem} \label{lem_axiom_disk}
Let $S$ be a closed oriented smooth 1-manifold. We have $\partial D_S = S$ with orientation. The assignment $S \mapsto D_s$ satisfies the following.

(a)(Functoriality) If $f : S \to S'$ is an orientation preserving diffeomorphism of closed oriented smooth 1-manifolds, then there exists a natural orientation preserving diffeomorphism $F : D_S \to D_{S'}$ such that $\partial F = f$.

(b)(Orientation) Let $-S$ be the manifold with the opposite orientation to $S$. There exists a natural identification $D_{-S} \cong - D_S$.
\end{lem}

\begin{proof}
If an orientation preserving diffeomorphism $f$ is given, then we have the orientation preserving diffeomorphism $f \times \mathrm{id} : S \times [0,1] \to S' \times [0,1]$. This diffeomorphism induces the orientation preserving diffeomorphism $F$. This proves (a). (b) is obvious from the definition of $D_S$.
\end{proof}

We extend the definition of the WZ term $e^{W_\Sigma(g)}$ to the case that $\Sigma$ has boundary and the restriction of $g$ to the boundary is the constant map whose values are the unit $e \in G$. We put $\G_{\Sigma,0} = \{ h \in \G_\Sigma | \partial h = e \}$.

\begin{dfn} \label{dfn_WZ_boundary_unit}
Let $\Sigma$ be a compact oriented smooth 2-manifold with boundary $\partial \Sigma = S_1 \sqcup \cdots \sqcup S_n$. For $h \in \G_{\Sigma,0}$ we define the WZ term by
\begin{eqnarray}
\exp W_\Sigma(h) = \exp W_{\Sigma \cup D} (h \cup e) ,
\end{eqnarray}
where $\Sigma \cup D$ is a closed 2-manifold obtained by patching disks $D = D_1 \sqcup \cdots \sqcup D_n$, and $h \cup e$ is the extension of $h$ on the closed 2-manifold by the unit.
\end{dfn}

\begin{lem}
For a compact oriented smooth 2-manifold $\Sigma$ with boundary, we define a function $\chi_\Sigma : \G_\Sigma \times \G_{\Sigma,0} \to \T$ by $\chi_\Sigma(f, h) := \exp \left\{ \Gamma_\Sigma (f, h) + W_\Sigma(h) \right\}$. This function satisfies the following formula
\begin{eqnarray}
\chi_\Sigma(f, h_1 h_2) = \chi_\Sigma(f h_1, h_2) \chi_\Sigma(f, h_1).
\end{eqnarray}
\end{lem}

\begin{proof}
We can easily check the formula
\begin{eqnarray}
\Gamma_\Sigma(h_1, h_2) + \Gamma_\Sigma(h_0, h_1 h_2) 
= \Gamma_\Sigma(h_0 h_1, h_2) + \Gamma_\Sigma(h_0, h_1) .
\label{cocycle_condition_2}
\end{eqnarray}
Using (\ref{cocycle_condition_2}) and (\ref{PW-formula_1}) we can prove this lemma.
\end{proof}

We define a Hermitian line bundle over $\G_S$ and the Wess-Zumino term in the case that $\Sigma$ has boundary by the similar way to the CS line bundle (Proposition \ref{thm_cslinebundle}) and the CS action (Proposition \ref{thm_csaction_boundary}).

\begin{dfn} \label{dfn_WZ_boundary}
(a) Let $S$ be a closed oriented smooth 1-manifold. We define a line bundle over $\G_S$ by putting
\begin{eqnarray}
K_S = \G_{D_S} \times \C / \sim ,
\end{eqnarray}
where the equivalence relation is $(f, z) \sim (fh, \chi_{-D_S}(f,h) z)$ for $h \in \G_{\Sigma,0}$. The projection $q_S : K_S \to \G_S$ is defined by $q_S([f, z]) := \partial f$ and the Hermitian structure is defined by the standard Hermitian structure on $\C$. We denote the fiber of the line bundle as $K_S(\gamma) := K_S|_\gamma$.

(b) Let $\Sigma$ be a compact oriented smooth 2-manifold with boundary $\partial \Sigma = S$. For $g \in \G_\Sigma$ we put $\gamma = \partial g$. We define the WZ term as an element of unit norm in $K_S(\gamma)$ by the following.
\begin{eqnarray}
e^{ W_\Sigma(g) } = [ f, e^{ W_{\Sigma \cup (-D_S)}(g \cup f) } ],
\end{eqnarray}
where $f \in \G_{-D_S} = \G_{D_S}$ is an element such that $\partial f = \gamma$.
\end{dfn}

\begin{rem}
If we identify $\C$ with $K_S(e)$, then Definition \ref{dfn_WZ_boundary_unit} is compatible with Definition \ref{dfn_WZ_boundary}.
\end{rem}

The Hermitian line bundle $K_S$ and the WZ term $e^{ W_\Sigma(g) }$ satisfy the following fundamental properties. The CS line bundle and the CS action satisfy similar properties \cite{F1}.

\begin{prop}[\cite{F1}] \label{thm_axioms_WZ}
Let $S$ be a closed oriented smooth 1-manifold and $\Sigma$ a compact oriented smooth 2-manifold with boundary. The assignments
\begin{eqnarray*}
\gamma \in \G_S & \longmapsto & K_S(\gamma) , \\
g \in \G_\Sigma & \longmapsto & e^{ W_\Sigma(g) } 
\in K_{\partial \Sigma}(\partial g)
\end{eqnarray*}
satisfy the following properties.

(a) (Functoriality) If $f : S' \to S$ be an orientation preserving diffeomorphism, then we have a natural isometry
\begin{eqnarray}
\h{\phi}_f : K_S(\gamma) \to K_{S'}(f^*\gamma) .
\end{eqnarray}
Moreover if there exists an orientation preserving diffeomorphism $F : \Sigma' \to \Sigma$ for 2-manifolds $\Sigma$ and $\Sigma'$ such that $\partial \Sigma = S, \partial \Sigma' = S'$ and $\partial F = f$, then we have
\begin{eqnarray}
\h{\phi}_f \left( e^{ W_\Sigma(g) } \right) = e^{ W_{\Sigma'}(F^*g) } .
\end{eqnarray}

(b) (Orientation) Let $-S$ be the manifold with the opposite orientation to $S$. Then we have a natural isomorphism
\begin{eqnarray}
K_{-S}(\gamma) \cong K_{S}^*(\gamma)
\hspace{1cm} \textrm{(the dual line bundle)}.
\end{eqnarray}
Moreover if $\partial \Sigma = S$ then we have
\begin{eqnarray}
e^{ W_{-\Sigma}(g) }  = \overline{ e^{ W_\Sigma(g) } }
\hspace{1cm} \textrm{(complex conjugation)}.
\end{eqnarray}

(c) (Additivity) If $S = S_1 \sqcup \cdots \sqcup S_n$ with $\gamma = \gamma_1 \sqcup \cdots \sqcup \gamma_n$, then we have a natural isomorphism
\begin{eqnarray}
K_S(\gamma) \cong 
K_{S_1}(\gamma_1) \otimes \cdots \otimes K_{S_n}(\gamma_n).
\end{eqnarray}
Moreover if $S$ is the boundary of $\Sigma = \Sigma_1 \sqcup \cdots \sqcup \Sigma_n$ with $g = g_1 \sqcup \cdots \sqcup g_n$, then we have
\begin{eqnarray}
e^{ W_\Sigma(g) } = 
e^{ W_{\Sigma_1}(g_1) } \otimes \cdots \otimes e^{ W_{\Sigma_n}(g_n) }
\end{eqnarray}
under the identification.

(d) (Gluing) Let $\Sigma$ be a 2-manifold, $S$ a closed 1-manifold, $j : S \to \Sigma$ an embedding, and $\Sigma_c$ the manifold obtained by cutting $\Sigma$ along $S$. For $g \in \G_\Sigma$ we denote the corresponding element in $\G_{\Sigma_c}$ by $g_c$. Using the contraction at $\gamma := j^*g$
\begin{eqnarray}
\Tr_\gamma : K_{\partial \Sigma_c}(\partial g_c) \cong 
K_{\partial \Sigma}(\partial g) \otimes K_S^*(\gamma) \otimes K_S(\gamma)
\rightarrow K_{\partial \Sigma}(\partial g) 
\end{eqnarray}
we have
\begin{eqnarray}
\Tr_\gamma e^{ W_{\Sigma_c}(g_c) } = e^{ W_\Sigma(g) } .
\end{eqnarray}
\end{prop}

\begin{proof}
These properties are the consequences of the properties of the disk $D_S$. (a) and (b) are proved directly using Lemma \ref{lem_axiom_disk}. (c) is proved by the definition of the disk $D_S$. We can prove (d) by writing explicitly the contraction of the line bundles $K_S^* \otimes K_S \cong \G_S \times \C$.
\end{proof}

Using the Hermitian line bundle $K_S$ over $\G_S$ we define the central extension of the loop group $\G_S$ based on Mickelsson's construction \cite{Mi}. The central extension is called the Kac-Moody group and is studied in detail \cite{P-S}. 

\begin{dfn}[\cite{Mi}] \label{dfn_KM_group}
Let $S$ be a closed oriented smooth 1-manifold. We define a group structure on
$\h{\G}_S = \{ \h{\gamma} \in K_S |\ |\h{\gamma}| = 1  \}$
as follows. Any element $\h{\gamma} \in \h{\G}_S$ is expressed as $\h{\gamma} = [g, u]$ for $g \in \G_{D_S}$ and $u \in \T$. For $\h{\gamma}_1 = [f_1, u_1]$ and $\h{\gamma}_2 = [f_2, u_2]$ we put
\begin{eqnarray*}
\h{\gamma_1} \cdot \h{\gamma_2} 
& = & 
[f_1, u_1] \cdot [f_2, u_2]  \\
& := &
[f_1 f_2, u_1 u_2 \exp \Gamma_{-D_S}(f_1, f_2)]
\end{eqnarray*}
The unit is $\hat{e} = [e, 1]$ and the inverse of $[f, u]$ is $[f, u]^{-1} = [f^{-1}, u^{-1}]$. The group $\h{\G}_S$ is a central extension of the loop group $\G_S$ by $\T$
$$
\begin{CD}
1 @>>> \T @>i_S>> \h{\G}_S @>q_S>> \G_S @>>>1 ,
\end{CD}
$$
where $i_S(u) = u \cdot \hat{e} = [e, u]$.
\end{dfn}

Note that the value of the WZ term formulated in Definition \ref{dfn_WZ_boundary} is an element of this group. We can easily check the following by (\ref{PW-formula_1}).

\begin{prop}
Let $\Sigma$ be a compact oriented smooth 2-manifold with boundary $\partial \Sigma = S$. As an element of $\h{\G}_S$ the WZ term $e^{ W_\Sigma (\cdot) }$ satisfies the Polyakov-Wiegmann formula
\begin{eqnarray}
e^{ W_\Sigma(g_1 g_2) } = 
\exp \Gamma_\Sigma(g_1,g_2) e^{ W_\Sigma(g_1) } \cdot e^{ W_\Sigma(g_2) } .
\label{PW-formula_2}
\end{eqnarray}
\end{prop}

The following is essential to the construction of the liftings in Section \ref{la_results}.

\begin{cor}
Let $Q$ be a smooth principal $G$-bundle over a compact oriented smooth 2-manifold $\Sigma$ with boundary. Using the WZ term of Definition \ref{dfn_WZ_boundary} we define a function $c_\Sigma : \A_Q \times \S_Q \times \G_\Sigma \to \h{\G}_{\partial \Sigma}$ by
\begin{eqnarray}
c_\Sigma(A,s,g) = 
\exp \frac{k\sqrt{-1}}{4\pi} \int_\Sigma \Tr(s^*A \wedge dgg^{-1}) \cdot \left( e^{W_\Sigma(g)} \right)^{-1} .
\label{cocycle_3}
\end{eqnarray}
Then $c_\Sigma$ satisfies the following formula
\begin{eqnarray}
c_\Sigma(A,s,g_1 g_2) = c_\Sigma(A,s \cdot g_1,g_2) \cdot c_\Sigma(A,s,g_1) ,
\label{cocycle_condition_3}
\end{eqnarray}
where the right hand side is the group product in $\h{\G}_{\partial \Sigma}$.
\end{cor}


\section{Introduction to $\T$-gerbes}
\label{la_intro_gerbe}

\textit{Gerbes} are invented by Giraud \cite{Gi}. In \cite{Br1}, Brylinski investigates \textit{Dixmier-Douady sheaves of groupoids} or \textit{DD gerbes} as geometric objects corresponding to degree 3 integral cohomology classes. The gerbe with which we deal in this paper is thought of as a ``fiber" of a DD gerbe. The formulations here are based on \cite{F2}. 

\begin{dfn} \label{dfn_torsor}
A manifold $T$ is called a \textit{$\T$-torsor}, or a \textit{principal $\T$-space}, if there exists a simply transitive (right) action $T \times \T \to T$. A morphism of principal $\T$-torsors is a map $f : T_1 \to T_2$ which commutes with the $\T$-actions. We denote the category of $\T$-torsors by $\sT$.
\end{dfn}

$\T$ itself is a $\T$-torsor and called the trivial $\T$-torsor. If a $\T$-bundle over $X$ is given, then the fiber at a point in $X$ is a $\T$-torsor. Note that each $\T$-torsor is isomorphic to the trivial $\T$-torsor $\T$, because a choice of a point on a $\T$-torsor gives an isomorphism. But, in general, we have no preferred point and there are no canonical isomorphisms between $\T$-torsors. Note also that all morphisms of $\T$-torsors are invertible. Hence the category $\sT$ is a \textit{groupoid}.

\begin{dfn}
(a) Let $T_1$ and $T_2$ be $\T$-torsors. We define an equivalence relation on the manifold $T_1 \times T_2$ by $\left( t_1 \cdot u, t_2 \right) \sim \left( t_1, t_2 \cdot u \right)$ for $u \in \T$. We denote the equivalence class of $(t_1, t_2)$ by $t_1 \otimes t_2$. The quotient space is a $\T$-torsor called the \textit{product} of $T_1$ and $T_2$, under the action $t_1 \otimes t_2 \mapsto t_1 \otimes (t_2 \cdot u)$. 

(b) For a $\T$-torsor $T$ an isomorphism $f : T \to \T$ of $\T$-torsors is called a \textit{trivialization} of $T$. We denote by $T^*$ the set of all trivializations of $T$. This manifold is a $\T$-torsor called the \textit{inverse} of $T$ under the action $f \mapsto f \cdot u$, where $f \cdot u : T \to \T$ is defined by $(f \cdot u)(t) := f(t) \cdot u$. 
\end{dfn}

The product and the inverse give $\sT$ an abelian group-like structure \cite{F2}. The structure is illustrated by the following proposition, which is easily proved.

\begin{prop} \label{prop_agls_torsors}
There exist natural isomorphisms of $\T$-torsors which commute with each other: 

(a) (associativity) $(T_1 \otimes T_2) \otimes T_3 \cong T_1 \otimes (T_2 \otimes T_3)$.

(b) (unit) $\T \otimes T \cong T \cong T \otimes \T$.

(c) (inverse) $T^* \otimes T \cong \T \cong T \otimes T^*$.

(d) (commutativity) $T_1 \otimes T_2 \cong T_2 \otimes T_1$.
\end{prop}

Now we introduce a $\T$-gerbe as a category with a ``simply transitive action of $\sT$." The precise treatment of categories takes too many pages, so we often omit details. The notation such as $P \in \sC$ means that $P$ is an object in a category $\sC$.

\begin{dfn}
Let $\sC$ be a category and $\varphi : \sC \times \sT \to \sC$ a functor. The functor is called a \textit{$\sT$-action} if it makes the following diagrams commutative up to natural equivalences.
$$
\begin{CD}
\sC \times \sT \times \sT @>{\Id \times \otimes}>> \sC \times \sT \\
@V{\varphi \times \Id}VV     @VV{\varphi}V  \\
\sC \times \sT @>>{\varphi}> \sC
\end{CD}
\hspace{2cm}
\begin{CD}
\sC @>{(\Id, e_\T)}>> \sC \times \sT \\
@|     @VV{\varphi}V  \\
\sC @>>{\Id}> \sC
\end{CD}
$$
In these diagrams $\otimes : \sT \times \sT \to \sT$ is the functor defined by the product of $\T$-torsors, and $e_\T : \sC \rightarrow \sT$ is the functor assigning all objects in $\sC$ to $\T$ and all morphisms to the identity $1 : \T \to \T$. We also require that $\varphi$ is compatible with the isomorphisms of Proposition \ref{prop_agls_torsors}. A $\sT$-action is \textit{simply transitive} if the functor $\sC \times \sT \to \sC \times \sC$ defined by $(P, T) \mapsto (P, \varphi(P, T))$ is an equivalence of categories. 
\end{dfn}

Because an object in a category may have inner structures, we have to distinguish equalities and isomorphisms strictly. For objects $P \in \sC$ and $T \in \sT$ we denote $\varphi(P, T) = P \cdot T$. The commutativity of two diagrams means that there are natural isomorphisms $P \cdot \T \cong P$ and $(P \cdot T_1) \cdot T_2 \cong P \cdot (T_1 \otimes T_2)$. The simply transitivity of a $\sT$-action amounts to the existence of a unique $\T$-torsor $T(P, Q)$ for $P, Q \in \sC$ together with a unique isomorphism $P \cdot T(P, Q) \cong Q$.

\begin{dfn}
A category $\sC$ is called a \textit{$\T$-gerbe} if it is a groupoid (i.e.\ all morphisms are invertible) and has a simply transitive $\sT$-action. An isomorphism of $\T$-gerbes $F : \sC_1 \to \sC_2$ is an equivalence of categories which makes the following diagram commutative up to a natural equivalence.
$$
\begin{CD}
\sC_1 \times \sT @>{\varphi_1}>> \sC_1 \\
@V{F \times \Id}VV  @VV{F}V \\
\sC_2 \times \sT @>{\varphi_2}>> \sC_2
\end{CD}
$$
\end{dfn}

\begin{rem}
We can consider a $\T$-gerbe as a ``DD gerbe \cite{Br1,Br2} on the space consisting of one point," because we can identify $\T$ with the group of automorphisms of an object in a $\T$-gerbe by means of the natural isomorphism $P \cdot \T \cong P$.
\end{rem}

If we define a $\sT$-action on $\sT$ itself by the product, then $\sT$ is a $\T$-gerbe and is called the trivial $\T$-gerbe. Note that each $\T$-gerbe are isomorphic to the trivial $\T$-gerbe $\sT$ as in the case of $\T$-torsors. $\T$-gerbes also have product and inverse. These operations give the ``category of $\T$-gerbes" an abelian group-like structure. In this paper, we distinguish the products of gerbes from that of torsors by using $\odot$.

\begin{dfn}
(a) For $\T$-gerbes $\sC_1$ and $\sC_2$ we define the \textit{product} $\sC_1 \odot \sC_2$ as follows. The objects in $\sC_1 \odot \sC_2$ are the same as $\sC_1 \times \sC_2$. The morphisms of $\sC_1 \odot \sC_2$ are that of $\sC_1 \times \sC_2$ and formal isomorphisms $(P_1, P_2) \to (P_1 \cdot T, P_2 \cdot T^*)$ for $T \in \sT$ and $(P_1, P_2) \in \sC_1 \times \sC_2$. The $\sT$-action is defined as $(P_1, P_2) \cdot T = (P_1, P_2 \cdot T)$. We denote the object in $\sC_1 \odot \sC_2$ corresponding to an object $(P_1, P_2)$ in $\sC_1 \times \sC_2$ by $P_1 \odot P_2$. 

(b) The \textit{inverse} $\sC^*$ of a $\T$-gerbe $\sC$ is defined as the following category. The objects are isomorphisms of $\T$-gerbes $F : \sC \to \sT$. The morphisms are natural equivalences between isomorphisms of $\T$-gerbes. The $\sT$-action is defined as $F \mapsto F \cdot T$, where $F \cdot T$ is the functor assigning an object $P$ to $F(P) \otimes T$.
\end{dfn}

\begin{prop}
There exist natural isomorphisms of $\sT$-gerbes which commute with each other: 

(a) (associativity) $(\sC_1 \odot \sC_2) \odot \sC_3 \cong \sC_1 \odot (\sC_2 \odot \sC_3)$.

(b) (unit) $\sT \odot \sC \cong \sC \cong \sC \odot \sT$.

(c) (inverse) $\sC^* \odot \sC \cong \sT \cong \sC \odot \sC^*$.

(d) (commutativity) $\sC_1 \odot \sC_2 \cong \sC_2 \odot \sC_1$.
\end{prop}

We denote the isomorphism of (c) by $\Tr$ and call the \textit{contraction}.

\begin{proof}
For (a), we construct the functor by $(P_1 \odot P_2) \odot P_3 \mapsto P_1 \odot (P_2 \odot P_3)$. For (d), we construct the functor by $P_1 \odot P_2 \mapsto P_2 \odot P_1$. By the definition of the product of $\T$-gerbes these functors give the isomorphism of $\T$-gerbes. For (b), the functor $\sC \odot \sT \to \sC$ is given by $P \odot T \mapsto P \cdot T$. For (c), the functor $\sC^* \odot \sC \to \sT$ is given by $F \odot P \mapsto F(P)$. Because $\sT$ acts on $\sC$ simply transitively, these functors are isomorphisms of $\T$-gerbes.
\end{proof}

We could define higher gerbes \cite{F2} by the inductive construction using the abelian group-like structures. But these are beyond the scope of this paper.


\section{Gerbes associated with liftings}
\label{la_lifting_gerbe}

We introduce the $\T$-gerbes given by liftings of torsors and relate the operations on liftings with that on gerbes. 

\smallskip

First, we generalize the $\T$-torsors by replacing $\T$ by a Lie group which may be non-abelian. When a manifold $\Xi$ admits a simply transitive (right) action of a Lie group $\Gamma$, we call $\Xi$ a \textit{$\Gamma$-torsor} (recall Definition \ref{dfn_torsor}). A typical example of a $\Gamma$-torsor is a fiber of a principal $\Gamma$-bundle.

\begin{dfn}
Let $\Gamma$ be a Lie group and $\h{\Gamma}$ a central extension $\h{\Gamma}$ by $\T$ 
$$
\begin{CD}
1 @>>> \T @>i>> \h{\Gamma} @>q>> \Gamma @>>> 1.
\end{CD}
$$
A \textit{lifting} of a $\Gamma$-torsor $\Xi$ is a pair $(\h{\Xi}, \hq)$, where $\h{\Xi}$ is a $\h{\Gamma}$-torsor and $\hq : \h{\Xi} \to \Xi$ is an equivariant map under the group actions: $\hq(\h{\xi} \cdot \h{\gamma}) = \hq(\h{\xi}) \cdot q(\h{\gamma})$. A morphism from $(\h{\Xi}, \hq)$ to $(\h{\Xi}', \hq')$ is a morphism $f : \h{\Xi} \to \h{\Xi}'$ as $\h{\Gamma}$-torsors which induces the identity morphism on $\Xi$.
\end{dfn}

We often abbreviate $(\h{\Xi}, \hq)$ as $\h{\Xi}$.

\begin{lem} \label{lem_dfn_sTaction}
Let $\Xi$ be a $\Gamma$-torsor and $\h{\Gamma}$ be a central extension of $\Gamma$. For a lifting $(\h{\Xi}, \hq)$ of $\Xi$ we put $\h{\Xi} \otimes T := (\h{\Xi} \times T)/\sim$, where the equivalence relation $\sim$ is defined by $(\h{\xi} \cdot u, t) \sim (\h{\xi}, t \cdot u)$ for $u \in \T$. We denote the equivalence class of $(\h{\xi}, t)$ as $\h{\xi} \otimes t$. We define a map $\hq' : \h{\Xi} \otimes T \to \Xi$ by $\hq'(\h{\xi} \otimes t) = \hq(\h{\xi})$. If we define a $\h{\Gamma}$-action on $\h{\Xi} \otimes T$ by $\h{\xi} \otimes t \mapsto (\h{\xi} \cdot \h{\gamma}) \otimes t$, then $(\h{\Xi} \otimes T, \hq')$ is a lifting of $\Xi$.
\end{lem}

\begin{proof}
Since $\T$ is the center of $\h{\Gamma}$, we can easily show this lemma.
\end{proof}

\begin{thm}
Let $\sT_{\h{\Gamma}}(\Xi)$ be the category of liftings of a $\Gamma$-torsor $\Xi$ for a central extension $\h{\Gamma}$ of $\Gamma$. If we define a $\sT$-action on $\sT_{\h{\Gamma}}(\Xi)$ by $(\h{\Xi}, \hq) \mapsto (\h{\Xi} \otimes T, \hq')$ using Lemma \ref{lem_dfn_sTaction}, then $\sT_{\h{\Gamma}}(\Xi)$ is a $\T$-gerbe.
\end{thm}

\begin{proof}
The map $\h{\Xi} \otimes \T \to \h{\Xi}$ defined by $\h{\xi} \otimes u \mapsto \h{\xi}u$ gives the isomorphism of the liftings. We can also construct the isomorphism $(\h{\Xi} \otimes T_1) \otimes T_2 \cong \h{\Xi} \otimes (T_1 \otimes T_2)$ easily. These isomorphisms give rise to a $\sT$-action on $\sT_{\h{\Gamma}}(\Xi)$. For liftings $\h{\Xi}_1, \h{\Xi}_2$ the set of all morphisms from $\h{\Xi}_1$ to $\h{\Xi}_2$ give rise to a $\T$-torsor which we denote by $T(\h{\Xi}_1, \h{\Xi}_2)$. There is a unique isomorphism $\h{\Xi}_1 \otimes T(\h{\Xi}_1, \h{\Xi}_2) \to \h{\Xi}_2$. Using this $\T$-torsor we can prove that the $\sT$-action is simply transitive.
\end{proof}

The following corollary will be used in Section \ref{la_results} to define the $\T$-gerbe assigned to a connection on a principal bundle over a 1-manifold.

\begin{cor} \label{cor_gerbe_bundle}
Let $p : B \to X$ be a principal $\Gamma$-bundle and $\h{\Gamma}$ be a central extension of $\Gamma$ by $\T$. For each $x \in X$ we have a $\T$-gerbe $\sT_{\h{\Gamma}}(B(x))$, where $B(x) := B|_x$ is the fiber of $B$ at $x$.
\end{cor}

\begin{rem}
The notion of lifting is usually defined for principal bundles. For a $\Gamma$-bundle $p : B \to X$ and a central extension $\h{\Gamma}$, a lifting $(\h{B}, \h{q})$ is a principal $\h{\Gamma}$-bundle $\h{p} : \h{B} \to X$ together with a map $\hq : \h{B} \to B$ satisfying $\hq(\h{b} \cdot \h{\gamma}) = \hq(\h{b}) \cdot q(\h{\gamma})$. Therefor the $\T$-gerbe $\sT_{\h{\Gamma}}(B(x))$ can be identified with the category of liftings of $\Gamma$-bundle $B(x) \to \{ x \}$.
\end{rem}

Since a central extension $\h{\Gamma}$ of a Lie group $\Gamma$ by $\T$ is a $\T$-bundle over $\Gamma$, there exist operations on central extensions corresponding to that on $\T$-bundles.

\begin{dfn} \label{dfn_operations_ce}
(a) (contracted product) Let $\h{\Gamma}_1$ and $\h{\Gamma}_2$ be central extensions of $\Gamma$. We denote the units by $\hat{e}_1$ and $\hat{e}_2$ respectively. We define a group product in the contracted product $\h{\Gamma}_1 \otimes \h{\Gamma}_2$ of $\T$-bundles by $(\h{\gamma}_1 \otimes \h{\gamma}_2) \cdot (\h{\eta}_1 \otimes \h{\eta}_2) := (\h{\gamma}_1 \cdot \h{\eta}_1) \otimes (\h{\gamma}_2 \cdot \h{\eta}_2)$. The unit is $\hat{e}_1 \otimes \hat{e}_2$. This group is a central extension
$$
\begin{CD}
1 @>>> \T @>i_1 \otimes i_2>> 
\h{\Gamma}_1 \otimes \h{\Gamma}_2 @>q_1 \otimes q_2>>  \Gamma @>>> 1, 
\end{CD}
$$
where $(i_1 \otimes i_2)(u) := u \h{e}_1 \otimes \h{e}_2$ and $(q_1 \otimes q_2)(\h{\gamma}_1 \otimes \h{\gamma}_2) := q_1(\h{\gamma}_1) = q_2(\h{\gamma}_2)$.

(b) (inverse) Let $\h{\Gamma}^*$ be the same manifold as $\h{\Gamma}$. We denote the element corresponding to $\h{\gamma} \in \h{\Gamma}$ by $\h{\gamma}^* \in \h{\Gamma}^*$. We define a group product in $\h{\Gamma}^*$ by $\h{\gamma}^*_1 \cdot \h{\gamma}^*_2 := ( \h{\gamma}_2 \cdot \h{\gamma}_1 )^*$. This group is a central extension
$$
\begin{CD}
1 @>>> \T @>i^*>> \h{\Gamma}^* @>q^*>> \Gamma @>>> 1,
\end{CD}
$$
where $i^*(u) := \overline{u} \h{e}^*$ ($\overline{u}$ denotes the complex conjugate of $u$) and $q^*(\h{\gamma}^*) := q(\h{\gamma})^{-1}$.

(c) (external product)
Let $\h{\Gamma}_1$ and $\h{\Gamma}_2$ be central extensions of $\Gamma_1$ and $\Gamma_2$ respectively. We denote the projections by $p_i : \Gamma_1 \times \Gamma_2 \to \Gamma_i$. The external product $\h{\Gamma}_1 \boxtimes \h{\Gamma}_2 := p_1^* \h{\Gamma}_1 \otimes p_2^* \h{\Gamma}_2$ of $\T$-bundles is a central extension
$$
\begin{CD}
1 @>>> \T @>\iota>> \h{\Gamma}_1 \boxtimes \h{\Gamma}_2 
@>q_1 \boxtimes q_2>> \Gamma_1 \times \Gamma_2 @>>> 1 ,
\end{CD}
$$
where $\iota(u) := u \h{e}_1 \boxtimes \h{e}_2$ and $(q_1 \boxtimes q_2) (\h{\gamma}_1 \boxtimes \h{\gamma}_2) := (q_1(\h{\gamma}_1), q_2(\h{\gamma}_2))$.
\end{dfn}

\begin{lem} \label{lem_tr_ce}
Let $\h{\Gamma}$ be a central extension of $\Gamma$, and $\h{\Gamma}^*$ the inverse of $\h{\Gamma}$. We define a homomorphism of groups
\begin{eqnarray*}
\Tr : \h{\Gamma}^* \otimes \h{\Gamma} \rightarrow \Gamma \times \T
\end{eqnarray*}
by $\Tr(\h{\gamma}^* \otimes \h{\eta}) := (q(\h{\eta}), \h{\gamma} \h{\eta})$. This is an isomorphism of central extensions.
\end{lem}

\begin{proof}
By the definition of the product we have $q^*(\h{\gamma}^*) = q(\h{\eta})$. So $\h{\gamma} \h{\eta}$ are expressed as an element of $\T$ and the map $\Tr$ is well-defined. The map is a homomorphism because $\T$ is the center. We can easily verify that the homomorphism is an isomorphism of groups which makes the following diagram commutative. 
$$
\begin{CD}
1 @>>> \T @>i^* \otimes i>> 
\h{\Gamma}^* \otimes \h{\Gamma} @>q^* \otimes q>>  \Gamma @>>> 1 \\
@. @| @V{\Tr}VV @| @. \\
1 @>>> \T @>>> \Gamma \times \T @>>> \Gamma @>>> 1
\end{CD}
$$
\end{proof}

These operations on central extensions give corresponding operations on liftings of $\Gamma$-torsors.

\begin{prop} \label{prop_operations_liftings}
(a)
Let $\Xi$ be a $\Gamma$-torsor, $(\h{\Xi}_1, \hq_1)$ and $(\h{\Xi}_2, \hq_2)$ liftings for central extensions $\h{\Gamma}_1$ and $\h{\Gamma}_2$ of $\Gamma$ respectively. On the contracted product $\hq_1 \otimes \hq_2 : \h{\Xi}_1 \otimes \h{\Xi}_2 \to \Xi$ as $\T$-bundles, we define a right action of $\h{\Gamma}_1 \otimes \h{\Gamma}_2$ by $(\h{\xi}_1 \otimes \h{\xi}_2) \cdot (\h{\gamma}_1 \otimes \h{\gamma}_2) := (\h{\xi}_1 \cdot \h{\gamma}_1) \otimes (\h{\xi}_2 \cdot \h{\gamma}_2)$. Then $(\h{\Xi}_1 \otimes \h{\Xi}_2, \hq_1 \otimes \hq_2)$ is a lifting for $\h{\Gamma}_1 \otimes \h{\Gamma}_2$.

(b)
Let $\Xi$ be a $\Gamma$-torsor and $(\h{\Xi}, \hq)$ a lifting for $\h{\Gamma}$. We define a right action of the inverse $\h{\Gamma}^*$ on $\h{\Xi}$ by $\h{\xi} \cdot \h{\gamma}^* := \h{\xi} \cdot \h{\gamma}^{-1}$. We denote the manifold $\h{\Xi}$ with the right action of $\h{\Gamma}^*$ by $\h{\Xi}^*$. If we define a map $\hq^* : \h{\Xi}^* \to \Xi$ by $\hq^* := \hq$, then we have a lifting $(\h{\Xi}^*, \hq^*)$ of $\Xi$ for $\h{\Gamma}^*$.

(c)
Let $\Xi_i$ be a $\Gamma_i$-torsor, $\h{\Gamma}_i$ a central extension of $\Gamma_i$, and $p_i : \Xi_1 \times \Xi_2 \to \Xi_i$ the projection $(i= 1, 2)$. For liftings $(\h{\Xi}_i, \hq_i)$ of $\Xi_i$ for $\h{\Gamma}_i$, the external product $\h{\Xi}_1 \boxtimes \h{\Xi}_2 := p_1^*\h{\Xi}_1 \otimes p_2^*\h{\Xi}_2$ as $\T$-bundles gives a lifting $(\h{\Xi}_1 \boxtimes \h{\Xi}_2, \hq_1 \boxtimes \hq_2)$ of the $\Gamma_1 \times \Gamma_2$-torsor $\Xi_1 \times \Xi_2$ for the central extension $\h{\Gamma}_1 \boxtimes \h{\Gamma}_2$.
\end{prop}

\begin{proof}
We can directly show this proposition using Definition \ref{dfn_operations_ce}.
\end{proof}

\begin{lem} \label{lem_tr_lifting}
Let $\h{\Gamma}$ be a central extension of $\Gamma$ and $(\h{\Xi}, \hq)$ be a lifting of a $\Gamma$-torsor $\Xi$ for $\h{\Gamma}$. Using the isomorphism of Lemma \ref{lem_tr_ce} we consider $\h{\Xi}^* \otimes \h{\Xi}$ as a $\Gamma \times \T$-torsor. There exists the isomorphism of liftings
\begin{eqnarray*}
\Tr : \h{\Xi}^* \otimes \h{\Xi} \rightarrow \Xi \times \T
\end{eqnarray*}
defined by $\Tr(\h{\xi}^* \otimes \h{\theta}) := (q(\h{\theta}), \zeta(\h{\xi}, \h{\theta}))$, where $\h{\xi} \cdot \zeta(\h{\xi}, \h{\theta}) = \h{\theta}$.
\end{lem}

\begin{proof}
We can show this lemma in the proof of Lemma \ref{lem_tr_ce}.
\end{proof}

We describe the relation between these operations on liftings and the abelian group-like structure of $\T$-gerbes, which is helpful to the proof of Theorem \ref{thm_axiom_CSgerbe}.

\begin{thm} \label{thm_gerbe_lifting}
Let $\Xi$ be a $\Gamma$-torsor. There exist natural isomorphisms of $\T$-gerbes

(a) $\sT_{\h{\Gamma}_1}(\Xi) \odot \sT_{\h{\Gamma}_2}(\Xi) \cong \sT_{\h{\Gamma}_1 \otimes \h{\Gamma}_2}(\Xi)$.

(b) $\sT_{\Gamma \times \T}(\Xi) \cong \sT$.

(c) $\sT_{\h{\Gamma}}(\Xi)^* \cong \sT_{\h{\Gamma}^*}(\Xi)$.
\end{thm}

\begin{proof}
For (a), let $\h{\Xi}_1$ and $\h{\Xi}_2$ be objects in $\sT_{\h{\Gamma}_1}(\Xi)$ and $\sT_{\h{\Gamma}_2}(\Xi)$ respectively. We define the functor $\sT_{\h{\Gamma}_1}(\Xi) \odot \sT_{\h{\Gamma}_2}(\Xi) \to \sT_{\h{\Gamma}_1 \otimes \h{\Gamma}_2}(\Xi)$ by assigning the formal object $\h{\Xi}_1 \odot \h{\Xi}_2$ to the product of liftings $\h{\Xi}_1 \otimes \h{\Xi}_2$. By the definition of $\sT$-action on $\sT_{\h{\Gamma}_1}(\Xi) \odot \sT_{\h{\Gamma}_2}(\Xi)$ the functor is well-defined.

The quasi-inverse functor of this functor is constructed as follows. Let $\h{\Xi}$ be a lifting of $\Xi$ for $\h{\Gamma}_1 \otimes \h{\Gamma}_2$. Take a trivialization $\h{\varphi} : \h{\Xi} \to \h{\Gamma}_1 \otimes \h{\Gamma}_2$. This induces the trivialization $\varphi : \Xi \to \Gamma$. We put $\h{\Xi}_i = \varphi^* \h{\Gamma}_i$ for $i = 1, 2$. The trivialization $\h{\varphi}$ induces the natural isomorphism of liftings $\h{\Xi} \cong \h{\Xi}_1 \otimes \h{\Xi}_2$. The lifting $\h{\Xi}_1 \otimes \h{\Xi}_2$ is unique up to the unique isomorphisms defined below. If $\h{\varphi}'$ is the other trivialization, there exists a unique element $\h{\gamma} \in \h{\Gamma}_1 \otimes \h{\Gamma}_2$ such that $\h{\gamma} \cdot \h{\varphi}(\h{\xi}) = \h{\varphi}'(\h{\xi})$ for all $\h{\xi} \in \h{\Xi}$. By multiplying $\h{\gamma}$ we have the unique isomorphism $f : \h{\Xi}_1 \otimes \h{\Xi}_2 \to \h{\Xi}'_1 \otimes \h{\Xi}'_2$. The unique isomorphism commute with the natural isomorphism from $\h{\Xi}$ to the product of liftings. Then the assignment $\h{\Xi} \mapsto \h{\Xi}_1 \odot \h{\Xi}_2$ gives the quasi-inverse functor.

For (b), let $\h{\Xi}$ be a lifting of $\Xi$ for the trivial central extension $\Gamma \times \T$. Since the quotient space $\h{\Xi}/\Gamma$ is a $\T$-torsor, we have a functor $\sT_{\Gamma \times \T}(\Xi) \to \sT$. If a $\T$-torsor $T$ is given, then $\Xi \times T$ is clearly a lifting of $\Xi$ for the trivial central extension. This gives the quasi-inverse functor.

For (c), by the help of Lemma \ref{lem_tr_lifting} and the isomorphism of (b), a lifting $\h{\Xi}^*$ for $\h{\Gamma}^*$ defines a functor $F_{\h{\Xi}^*} : \sT_{\h{\Gamma}}(\Xi) \to \sT$. The assignment $\h{\Xi}^* \mapsto F_{\h{\Xi}^*}$ gives the isomorphism of $\T$-gerbes $\sT_{\h{\Gamma}^*}(\Xi) \to \sT_{\h{\Gamma}}(\Xi)^*$.
\end{proof}

\begin{cor} \label{cor_external_gerbe}
Let $\Xi$ be a $\Gamma_i$-torsor and $\h{\Gamma}_i$ a central extension of $\Gamma_i$ for $i = 1, 2$. We have a natural isomorphism of $\T$-gerbes
$$
\sT_{\h{\Gamma}_1}(\Xi_1) \odot \sT_{\h{\Gamma}_1}(\Xi_2) \cong 
\sT_{\h{\Gamma}_1 \boxtimes \h{\Gamma}_2}(\Xi_1 \times \Xi_2).
$$
\end{cor}

\begin{proof}
By using Proposition \ref{prop_operations_liftings} (c) the assignment $\h{\Xi}_1 \odot \h{\Xi}_2 \mapsto \h{\Xi}_1 \boxtimes \h{\Xi}_2$ defines the functor above. We construct a quasi-inverse functor of this functor as follows. Let $\h{\Xi}$ be a lifting of $\Xi_1 \times \Xi_2$ for the central extension $\h{\Gamma}_1 \boxtimes \h{\Gamma}_2 := p_1^*\h{\Gamma}_1 \otimes p_2^*\h{\Gamma}_2$. By the help of Theorem \ref{thm_gerbe_lifting} (a) we can write $\h{\Xi} \cong \tilde{\Xi}_1 \otimes \tilde{\Xi}_2$, where $\tilde{\Xi}_i$ is a lifting of $\Xi_1 \times \Xi_2$ for the central extension $p_i^*\h{\Gamma}_i$ $(i = 1, 2)$. Since $p_1^*\h{\Gamma}_1 \cong \h{\Gamma}_1 \times \Gamma_2$, we obtain the lifting $\tilde{\Xi}_1 / \Gamma_2$  of $\Xi_1$ for $\h{\Gamma}_1$. Similarly we obtain the lifting $\tilde{\Xi}_2 / \Gamma_1$  of $\Xi_2$ for $\h{\Gamma}_2$. Then the assignment $\h{\Xi} \mapsto (\tilde{\Xi}_1 / \Gamma_2) \odot (\tilde{\Xi}_2 / \Gamma_1)$ gives the quasi-inverse functor.
\end{proof}


\section{Main results}
\label{la_results}

Here we state the main results of this paper. We fix a Lie group $G = SU(2)$.

\begin{dfn}
Let $S$ be a closed oriented smooth 1-manifold and $R \to S$ a smooth principal $G$-bundle. We have a central extension $\h{\G}_S$ of $\G_S$ and a $\G_S$-bundle $B_R = \A_R \times \S_R$ over $\A_R$, where the loop group acts only on the space $\S_R$ of smooth sections of $R$. Using Corollary \ref{cor_gerbe_bundle} We define a $\T$-gerbe assigned to $a \in \A_R$ by ${\B}_{(R, S)}(a) := \sT_{\h{\G}_S}(B_R(a))$.
\end{dfn}

\begin{thm}
Let $Q$ be a smooth principal $G$-bundle over a compact oriented smooth 2-manifold $\Sigma$ with boundary. For a connection $A$ on $Q$ we define a quotient space 
\begin{eqnarray}
\h{B}_{(Q, \Sigma)}(A) = 
\{A\} \times \S_Q \times \h{\G}_{\partial \Sigma} / \sim
\end{eqnarray} 
by the equivalence relation $(A, s, \h{\gamma}) \sim (A, s \cdot g, c_\Sigma(A,s,g) \cdot \h{\gamma})$ for $g \in \G_\Sigma$, where $c_\Sigma$ is defined by (\ref{cocycle_3}). If we define a right action by $[A, s,\h{\gamma}] \cdot \h{\eta} = [A, s,\h{\gamma} \cdot \h{\eta}]$ and a map $\hq_{(Q, \Sigma)} : \h{B}_{(Q, \Sigma)}(A)  \to  B_{\partial Q}(\partial A)$ by $\hq_{(Q, \Sigma)}([A, s, \h{\gamma}]) = (\partial A, \partial s \cdot q_S(\h{\gamma}))$, then $(\h{B}_{(Q,\Sigma)}, \hq_{(Q, \Sigma)})$ is a lifting of $B_{\partial Q}(\partial A)$, i.e.\ an object in $\B_{(\partial Q, \partial \Sigma)}(\partial A)$.
\end{thm}

\begin{proof}
The relation $\sim$ is indeed an equivalence relation because of (\ref{cocycle_condition_3}). Clearly the right action is well-defined and is simply transitive. The map $\hq_{(Q, \Sigma)}$ is also well-defined since $q_S(c_\Sigma(A, s, g)) = \partial g^{-1}$. We can easily check that it commutes with the actions of the groups.
\end{proof}

When the underlying manifolds are obvious, we abbreviate the notations as $\B_{(R, S)} = \B_R$ and $(\h{B}_{(Q, \Sigma)}, \hq_{(Q, \Sigma)}) = (\h{B}_Q, \hq_Q)$.

\begin{rem}
The construction above gives directly a lifting $(\h{B}_Q, \hq_Q)$ on the $\G_{\partial \Sigma}$-bundle $r^*B_{\partial Q} \to \A_Q$, where $r : \A_Q \to \A_{\partial Q}$ is the restriction map. We can consider the lifting as a $\T$-bundle over $r^*B_{\partial Q} = \A_Q \times \S_{\partial Q}$. The Hermitian line bundle associated with this $\T$-bundle corresponds to the Chern-Simons line bundle with $\partial \Sigma \neq \emptyset$ defined by Freed in \cite{F1}. 
\end{rem}

\begin{thm} \label{thm_axiom_CSgerbe}
Let $R \to S$ be a $G$-bundle over a closed oriented smooth 1-manifold, and $Q \to \Sigma$ be a $G$-bundle over a compact oriented smooth 2-manifold with boundary. The assignments
\begin{eqnarray*}
a \in \A_R & \longmapsto & \B_R(a), \\
A \in \A_Q & \longmapsto & (\h{B}_Q(A), \hq_Q) \in \B_{\partial Q}(\partial A)
\end{eqnarray*}
satisfy the following properties.

(a) (Functoriality) If $f : R' \to R$ be a bundle map covering an orientation preserving diffeomorphism of 1-manifolds, then we have a natural isomorphism of $\T$-gerbes
\begin{eqnarray}
\h{\Psi}_f : \B_R(a) \rightarrow \B_{R'}(f^*a) .
\label{axiom_CS_a1}
\end{eqnarray}
Moreover if there are compact 2-manifolds $Q$, $Q'$ such that $\partial Q = R, \partial Q' = R'$ and a bundle map $F : Q' \to Q$ covering an orientation preserving diffeomorphism such that $\partial F = f$, then we have a natural isomorphism of liftings
\begin{eqnarray}
\h{\Psi}_f \left( (\h{B}_Q(A), \hq_Q) \right) \cong 
(\h{B}_{Q'}(F^*A), \hq_{Q'}).
\end{eqnarray}

(b) (Orientation) Let $-S$ be the manifold with the opposite orientation to $S$. Then we have a natural isomorphism of $\T$-gerbes
\begin{eqnarray}
\B_{(R, -S)}(a) \cong \B_{(R, S)}(a)^* .
\label{axiom_CS_b1}
\end{eqnarray}
Moreover if $\partial \Sigma = S$ then we have a natural isomorphism of liftings
\begin{eqnarray}
( \h{B}_{(Q, -\Sigma)}(A),  \hq_{(Q, -\Sigma)}  ) \cong 
( \h{B}_{(Q, \Sigma)}(A)^*, \hq_{(Q, \Sigma)}^* ) .
\label{axiom_CS_b2}
\end{eqnarray}

(c) (Additivity) If $R = R_1 \sqcup \cdots \sqcup R_n$ with $a = a_1 \sqcup \cdots \sqcup a_n$, then we have a natural isomorphism of $\T$-gerbes
\begin{eqnarray}
\B_R(a) \cong 
\B_{R_1}(a_1) \odot \cdots \odot \B_{R_n}(a_n). 
\label{axiom_CS_c1}
\end{eqnarray}
Moreover if $R$ is the boundary of $Q = Q_1 \sqcup \cdots \sqcup Q_n$ with $A = A_1 \sqcup \cdots \sqcup A_n$, then we have a natural isomorphism of liftings
\begin{eqnarray}
(\h{B}_Q(A), \hq_Q) 
& \cong &
(\h{B}_{Q_1}(A_1), \hq_{Q_1}) \odot \cdots \odot 
(\h{B}_{Q_n}(A_n), \hq_{Q_n}) .
\label{axiom_CS_c2}
\end{eqnarray}

(d) (Gluing) Let $\Sigma$ be a 2-manifold, $S$ a closed 1-manifold, $j : S \to \Sigma$ an embedding, and $\Sigma_c$ the manifold obtained by cutting $\Sigma$ along $S$. For a $G$-bundle $Q$ over $\Sigma$ we have the induced $G$-bundle $Q_c$ over $\Sigma_c$ and $R := j^*Q$ over $S$. For a connection $A \in \A_Q$ we denote the corresponding connections by $A_c \in \A_{Q_c}$ and $a := j^*A \in \A_R$. We have a natural isomorphism of liftings
\begin{eqnarray}
\Tr_a \left( (\h{B}_{Q_c}(A_c), \hq_{Q_c}) \right)
\cong (\h{B}_Q(A), \hq_Q) ,
\label{axiom_CS_d1}
\end{eqnarray}
where $\Tr_a$ is the contraction of $\T$-gerbes
\begin{eqnarray}
\Tr_a : \B_{\partial Q_c}(\partial A_c) \cong 
\B_{\partial Q}(\partial A) \odot \B_R(a)^* \odot \B_R(a) 
\rightarrow \B_{\partial Q}(\partial A) .
\end{eqnarray}
\end{thm}

\begin{proof}
For (a), $\overline{f} : S' \to S$ denotes the diffeomorphism covered by $f$. Let $(\h{B}, \hq)$ be a lifting of the $\G_S$-bundle $B_R(a)$. We define a $\h{\G}_{S'}$-action on $\h{B}$ via the isomorphism $\h{\phi}_{\overline{f}} : \h{\G}_S \to \h{\G}_{S'}$ induced from Proposition \ref{thm_axioms_WZ} (a). If we define a map $\Psi_f : B_R(a) \to B_{R'}(f^*a)$ by $\Psi_f(a, s) := (f^*a, f^{-1} \circ s \circ \overline{f})$, then $(\h{B}, \Psi_f \circ \hq)$ is a lifting of the $\G_{S'}$-bundle $B_{R'}(f^*a)$. This assignment of liftings gives rise to the isomorphism (\ref{axiom_CS_a1}). If a bundle map $F : Q' \to Q$ is given, then the map $\S_Q \times \h{\G}_{\Sigma} \to \S_{Q'} \times \h{\G}_{\Sigma'}$ defined by $(s, \h{\gamma}) \mapsto (F^{-1} \circ s \circ \overline{F}, \h{\phi}_{\overline{f}}(\h{\gamma}))$ induces the map $\h{B}_Q(A) \to \h{B}_{Q'}(F^*A)$. By the functoriality of the integration over manifolds the induced map gives the isomorphism of the liftings of $B_{\partial Q'}(f^* a)$
$$
\h{\Psi}_f \left( (\h{B}_Q(A), \hq_Q) \right) = 
(\h{B}_Q(A), \Psi_f \circ \hq_Q) \longrightarrow
(\h{B}_{Q'}(F^*A), \hq_{Q'}) .
$$

For (b), we have the isomorphism $\h{\G}_{-S} \cong \h{\G}^*$ of the central extensions of $\G_{-S} = \G_{S}$ from Proposition \ref{thm_axioms_WZ} (b). By the help of Theorem \ref{thm_gerbe_lifting} (c) we have the isomorphism of (\ref{axiom_CS_b1}). Since the integration over manifolds satisfies $\int_{-X} = -\int_X$, we can easily construct the isomorphism (\ref{axiom_CS_b2}).

For (c), we prove the case of $n = 2$. It is clear that $\G_{S_1 \sqcup S_2} = \G_{S_1} \times \G_{S_2}$. By using Proposition \ref{thm_axioms_WZ} (c) we get the isomorphism $\h{\G}_{S_1 \sqcup S_2} \cong \h{\G}_{S_1} \boxtimes \h{\G}_{S_2}$. We can easily see that $B_R(a) = B_{R_1}(a_1) \times B_{R_2}(a_2)$. Applying Corollary \ref{cor_external_gerbe} we have the isomorphism (\ref{axiom_CS_c1}). We also have the isomorphism (\ref{axiom_CS_c2}) using the additivity of the integration over manifolds.

For (d), we describe the image under the isomorphism $\Tr_a$ first. Let $\h{B}$ be a lifting of $\G_{\Sigma_c}$-bundle $B_{\partial Q_c}(\partial A_c)$. By choosing a trivialization of $\h{B}$ we write $\h{B} \cong \h{B}_1 \boxtimes \h{B}_2 \boxtimes \h{B}_3$, where $\h{B}_1, \h{B}_2$ and $\h{B}_3$ are liftings of $B_{\partial Q}(\partial A), B_{(R, -S)}(a)$ and $B_{(R, S)}(a)$ respectively. Using (b) we identify $\h{B}_2$ with a lifting on $B_{(R, S)}$ for the central extension $\G_S^*$. Under the isomorphism $\B_R(a)^* \odot \B_R(a) \cong \sT$ we obtain the corresponding $\T$-torsor $T = \h{B}_2 \otimes \h{B}_3 / \G_S$. Therefore we have $\Tr_a(\h{B}) = \h{B}_1 \otimes T$. This construction seems to depend on the choice of the trivialization. Recall the proof of Theorem \ref{thm_gerbe_lifting} (a). If we take the other trivialization, then we have the unique isomorphism $f : \h{B}_1 \boxtimes \h{B}_2 \boxtimes \h{B}_3 \to \h{B}'_1 \boxtimes \h{B}'_2 \boxtimes \h{B}'_3$. This induces the unique isomorphism $\Tr_a(f) : \h{B}_1 \to \h{B}'_1$. Hence the image of $\Tr_a$ is uniquely determined up to the unique isomorphism.

Now we put $\h{B} = \h{B}_{Q_c}(A_c)$. The trivialization is induced by choosing a section $s_{Q_c} \in \S_{Q_c}$. We put $\partial s_{Q_c} := (s_{\partial Q}, s_R, s_R') \in \S_{\partial Q_c} = \S_{\partial Q} \times \S_R \times \S_R$.  Though $s_R \neq s_R'$ in general, we can choose a section such that $s_R = s_R'$ by using the action of $\G_{\Sigma_c}$. We fix such a section $\sigma_{Q_c}$ and put $\partial \sigma_{Q_c} := (\sigma_{\partial Q}, \sigma_R, \sigma_R)$. Under this choice we can put $\h{B}_1 := \G_{\partial \Sigma}, \h{B}_2 := \G_{-S}$ and $\h{B}_3 := \G_S$, where the equivariant maps are defined by
$$
\begin{array}{ccc}
\h{B}_1 \rightarrow B_{\partial Q}(\partial A) , &
\h{B}_2 \rightarrow B_{(R, -S)}(a),  &
\h{B}_3 \rightarrow B_{(R,  S)}(a), \\
(\h{\gamma}_{\partial \Sigma} \mapsto 
\sigma_{\partial Q} \cdot 
q_{\partial \Sigma}(\h{\gamma}_{\partial \Sigma})) &
(\h{\gamma}_{-S} \mapsto \sigma_R \cdot q_{-S}(\h{\gamma}_{-S})) &
(\h{\gamma}_{ S} \mapsto \sigma_R \cdot q_{ S}(\h{\gamma}_{ S})) 
\end{array}
$$
respectively. We can easily see that the inverse lifting $\h{B}_2^*$ is naturally isomorphic to $\h{B}_3$. Hence we have natural isomorphisms $T \cong \T$ and $\h{B}_1 \otimes T \cong \h{B}_1$. If we take the other section $\sigma_{Q_c}'$, then there exists a unique element $g_c \in \G_{\Sigma_c}$ such that $\sigma_{Q_c}' = \sigma_{Q_c} \cdot g_c$. The isomorphism $f : \h{B}_1 \boxtimes \h{B}_2 \boxtimes \h{B}_3 \to \h{B}_1' \boxtimes \h{B}_2' \boxtimes \h{B}_3'$ is defined by $\h{\gamma}_{\partial \Sigma} \boxtimes \h{\gamma}_{-S} \boxtimes \h{\gamma}_S \mapsto c_{\Sigma_c}(A_c, \sigma_{Q_c}, g_c) \cdot \h{\gamma}_{\partial \Sigma} \boxtimes \h{\gamma}_{-S} \boxtimes \h{\gamma}_S$. Note that $\sigma_{Q_c}$ and $g_c$ give the corresponding elements $\sigma_Q \in \S_Q$ and $g \in \G_\Sigma$. Using Proposition \ref{thm_axioms_WZ} (d) and the property of the integration, we have $\Tr( c_{\Sigma_c}(A_c, \sigma_{Q_c}, g_c)) = c_\Sigma(A, \sigma_Q, g)$. Hence the unique isomorphism $\Tr_a(f)$ is defined by $\h{\gamma}_\Sigma \mapsto c_\Sigma(A, \sigma_Q, g) \cdot \h{\gamma}_\Sigma$. Therefore the isomorphism of liftings $\h{B}_1 \to \h{B}_Q(A)$ defined by $\h{\gamma}_\Sigma \mapsto [A, \sigma_Q, \h{\gamma}_\Sigma]$ gives rise to the isomorphism (\ref{axiom_CS_d1}). 
\end{proof}

\begin{cor}
If $\partial \Sigma = \emptyset$ in Theorem \ref{thm_axiom_CSgerbe} (d), we have a natural isomorphism of $\T$-torsors
\begin{eqnarray}
\Tr_a \left( (\h{B}_{Q_c}(A_c), \hq_{Q_c}) \right) \cong T_Q(A),
\end{eqnarray}
where $\Tr_a$ is the isomorphism of $\T$-gerbes
\begin{eqnarray}
\Tr_a : \B_{\partial Q_c}(\partial A_c) \cong \B_R(a)^* \odot \B_R(a) 
\rightarrow \sT .
\end{eqnarray}
\end{cor}

\bigskip

As the $\T$-torsor $T_Q(a)$ is related to the Hermitian line bundle on the space of connections $\A_Q$, the $\T$-gerbe $\B_R(a)$ is related to a certain geometric object on $\A_R$. That is a \textit{Dixmier-Douady sheaf of groupoids} or a \textit{DD gerbe} \cite{Br1, Br2}. We can define a natural DD gerbe on $\A_R$ by extending the construction of $\B_R(a)$. The precise definition and the differential geometry of the DD gerbes are the subjects of the forthcoming paper \cite{Go}.

\bigskip

\textit{Acknowledgments}. I warmly thank T. Kohno for many useful suggestions and discussions.



Graduate school of Mathematical Sciences, 
University of Tokyo, Komaba 3-8-1, Meguro-Ku, Tokyo, 153-8914 Japan.

e-mail: kgomi@ms.u-tokyo.ac.jp

\end{document}